\documentclass[3p]{elsarticle}  
\usepackage{amsfonts,amssymb, mathtools, amsmath,bm}

\begin{document}

\begin{frontmatter}

\title{Hamiltonian structure of reduced fluid models for plasmas obtained from a kinetic description}

\author{L. de Guillebon}
\ead{loic.de-guillebon@cpt.univ-mrs.fr}
\author{C. Chandre}
\ead{chandre@cpt.univ-mrs.fr}  
\address{Centre de Physique Th\'eorique, CNRS -- Aix-Marseille Universit\'e, Campus de Luminy, case 907, F-13288 Marseille cedex 09, France}
 
\begin{abstract}
We consider the Hamiltonian structure of reduced fluid models obtained from a kinetic description of collisionless plasmas by Vlasov-Maxwell equations. We investigate the possibility of finding Poisson subalgebras associated with fluid models starting from the Vlasov-Maxwell Poisson algebra. In this way, we show that the only possible Poisson subalgebra involves the moments of zeroth and first order of the Vlasov distribution, meaning the fluid density and the fluid velocity. 
We find that the bracket derived in [Phys. Rev. Lett. {\bf 93}, 175002 (2004)] which involves moments of order 2 is not a Poisson bracket since it does not satisfy the Jacobi identity.   
\end{abstract}
\end{frontmatter}

\section{Introduction}

In the kinetic framework, the dynamics of collisionless plasmas is provided by the Vlasov-Maxwell equations which are dynamical equations for the phase space density $f({\bf x},{\bf v})$ of the charged particles (also called Vlasov density) and the electromagnetic fields ${\bf E}({\bf x})$ and ${\bf B}({\bf x})$ where ${\bf x}$ and ${\bf v}$ belong to ${\mathbb R}^3$. Here, we work with only one species of charged particles of unit mass and charge $e$ for the sake of simplicity, but the generalization to several species is straightforward (see, e.g., Ref.~\cite{Spen82}). The dynamical equations are 
\begin{eqnarray}
&& \dot f = -{\mathbf{v}}\cdot{\mathbf{\nabla}} f - e({\mathbf{E}} + {\mathbf{v}}\times\mathbf{B})\cdot\partial_{\mathbf{v}} f, \label{eq:VM1}\\
&& \dot {\mathbf{E}} =  {\mathbf{\nabla}}\times\mathbf{B} - e\int d^3v f{\mathbf{v}}, \label{eq:VM2}\\
&& \dot {\mathbf{B}} = -{\mathbf{\nabla}}\times{\mathbf{E}}, \label{eq:VM3}
\end{eqnarray}
where the dot designates the time derivative. Given nowadays computer capability, integrating numerically this kinetic model is too demanding for realistic laboratory plasmas. Moreover the dynamics is not easily analyzed in such a kinetic framework since it provides dynamical information at temporal and spatial scales which might be irrelevant. In addition, the plasma dynamics is more conveniently analyzed in the configuration space rather than in the phase space of the charged particles. As a consequence, there is a need for reduction in order to eliminate the irrelevant parts of these equations so as to obtain a much more tractable model. One way of doing this is to consider a fluid reduction which is obtained by considering the moments of the Vlasov distribution, e.g., the fluid density, the fluid momentum, the pressure tensor, etc...\ 

Starting from the dynamics of the zeroth order moment, a series of dynamical equations for the higher order moments is built. This is a priori an infinite set of differential equations. The reduction is obtained by truncating this set and closing the system of equations through a closure assumption, often based on a collisional argument or on a local thermodynamical equilibrium. According to the physical situation of interest, the choice of an order of truncation and of a closure assumption brings us a corresponding fluid model. For instance, starting from the parent model, i.e., the Vlasov-Maxwell equations, many physically interesting fluid models have been derived (see, e.g., Refs.~\cite{Bitt04, Nich83, Morr98, Pass05, Shad11}). A particularly important property of the parent model is that it possesses a Hamiltonian structure, i.e., the equations of motion can be rewritten using a Hamiltonian functional $H$ and a Poisson bracket $\{\cdot ,\cdot\}$ as $\dot{F}=\{F,H\}$. Truncating and closing the set of equations might not conserve this property. This introduces dissipative or non-Hamiltonian terms in the equations of motion whether they have a physical origin (as described by phenomenological constants like diffusion constant, magnetic diffusivity or kinematic viscosity) or not. If these terms do not have a physical origin, they have been coined mutilations in Ref.~\cite{Morr05}. If one considers only the ideal part of the equations which is the part where the phenomenological constants characterizing the dissipative terms are set to zero, one should recover a Hamiltonian system as an inheritance of the Hamiltonian parent model~\cite{Morr98,Morr05,Morr06,Chan10,Chan12}. 

There are two ways to proceed. The first one is to work with the equations of motion and check a posteriori that the resulting set of reduced equations possesses a Hamiltonian structure (by finding an appropriate conserved quantity and a Poisson bracket). This can be tedious as one needs to check that the bracket which has been guessed satisfies all the properties of a Poisson bracket, and in particular the Jacobi identity. The second method is to work directly on the Hamiltonian structure of the parent model by performing the reduction on the Hamiltonian and on the bracket. Of course the main difficulty resides in verifying that the reduced bracket still satisfies the Jacobi identity. This puts some restrictions on what is allowed to do on a Poisson bracket in the course of the reduction. 
The gain is significant since the preservation of the Hamiltonian structure is ensured, i.e., there is no fake dissipation or mutilation, and it allows to keep track of the conserved quantities throughout the derivation and apply all the techniques already available for Hamiltonian systems (like perturbation theory, energy-Casimir methods for equilibria, Lie transforms, etc...).

In Sec.~\ref{sec:2}, we consider a very practical and useful method for deriving reduced fluid models by using Poisson subalgebra. We apply it to the Vlasov-Maxwell Hamiltonian structure. This method provides the Hamiltonian structure of the usual fluid model composed by the continuity equation for the fluid density and the momentum equation for the fluid velocity. However we show in Sec.~\ref{sec:3} that when the closure occurs at orders higher than one, the reduced structure is no longer a Poisson subalgebra associated with the parent structure. As a consequence it is not possible to obtain reduced fluid models containing higher order moments by just applying a Poisson subalgebra argument. 

In Refs.~\cite{Shad04, Shad11}, a fluid model for low temperature relativistic plasmas, called the warm fluid model, involving second order moments of the Vlasov distribution was proposed. It aims at investigating the interaction between a strong laser pulse and a low-density plasma. In Ref.~\cite{Shad04} a conserved quantity has been constructed based on the Vlasov-Maxwell Hamiltonian, and a bracket was proposed. It has been shown that this model conserves energy and entropy. However, based on the results explained in Sec.~\ref{sec:2}, the Hamiltonian property of this model has to be scrutinized since it involves higher order moments of the Vlasov distribution. In Sec.~\ref{sec:4} we exhibit a counter example for the Jacobi identity for the bracket proposed in Ref.~\cite{Shad04}. Therefore, this warm fluid model is not a Hamiltonian system.

\section{Reduction to the usual fluid model}
\label{sec:2}

The starting point is the Hamiltonian structure of the Vlasov-Maxwell equations~\cite{Morr80,Wein81,Mars82,Morr81}. The Hamiltonian functional is the total energy~:
$$
H(f,{\mathbf{E}},\mathbf{B}) =\int d^3v d^3x~ f~\frac{{\mathbf{v}}^2}{2}
		+ \int d^3x~ \frac{{\mathbf{E}}^2 + \mathbf{B}^2}{2}.
$$
The Poisson bracket acts on the Poisson algebra of observables, that is the set of functionals of the field variables $f({\mathbf{x}},{\mathbf{v}})$, ${\mathbf{E}}({\mathbf{x}})$ and $\mathbf{B}({\mathbf{x}})$~:
\begin{eqnarray}
\{F,G\}_V &=&
		\int d^3x d^3v~ f  [F_f , G_f] + e\int d^3x d^3v~ f \left( \partial_{\mathbf{v}} F_f \cdot G_{\mathbf{E}}- F_{\mathbf{E}} \cdot \partial_{\mathbf{v}} G_f\right)\nonumber \\
		&& \quad + \int d^3x~ \left( F_{\mathbf{E}} \cdot {\mathbf{\nabla}} \times G_{\mathbf{B}} -F_{\mathbf{B}} \cdot \nabla \times G_{\mathbf{E}}\right),
\label{VlasMaxw}
\end{eqnarray}
where $F_\psi$ indicates the functional derivative with respect to the field variable $\psi$ and the bracket $[\cdot,\cdot]$ is given by
$$
[f,g]={\mathbf{\nabla}} f\cdot \partial_{\mathbf{v}} g -\partial_{\mathbf{v}} f \cdot {\mathbf{\nabla}} g +e{\mathbf{B}} \cdot \partial_{\mathbf{v}} f\times \partial_{\mathbf{v}} g,
$$ 
and $\partial_{\bf v}$ designates the partial derivatives with respect to ${\bf v}$. 
The dynamics of a functional $F$ of the Poisson algebra is given by $\dot{F}=\{F,H\}$. In particular we recover Eqs.~(\ref{eq:VM1})-(\ref{eq:VM3}) if we choose $F=f$, $F={\bf E}$ or $F={\bf B}$. We notice that we have considered here a non-relativistic plasma. However the discussion which follows is unchanged in the case of a relativistic plasma since the changes only occur in the Hamiltonian, not in the Poisson bracket~\cite{Bial84}.  

Fluid models rely on the idea of replacing the Vlasov density $f$ by the series of its moments~\cite{Zakh97}
\begin{equation}
\label{DefMom}
	P_n^{i_1\ldots i_n} = \int d^3 v  f v_{i_1}\cdots v_{i_n},
\end{equation}
for $n\in {\mathbb N}$. 
The chain rule yields
$$
	F_f = \sum_{n=0}^\infty F_{{P_n}^{i_1 i_2\ldots i_n}} ~ v_{i_1} v_{i_2}\cdots v_{i_n} ,
$$
where we have used Einstein's convention of implicit summation over repeated indices $i_k$. The expression of the Poisson bracket in these new variables involves the derivatives with respect to ${\bf v}$ given by
$$
\partial_{v_l} F_f=\sum_{n=0}^\infty \sum_{k=1}^n F_{P_n^{i_1 \ldots i_n}} ~ v_{i_1}\cdots v_{i_{k-1}} \delta_{i_k}^l v_{i_{k+1}}\cdots v_{i_n}.
$$
We use the following symmetrization of the tensor $P_n$
$$
F_{P_n^{(i_1 \ldots i_n)}}=\frac{1}{n}\sum_{k=0}^{n-1} F_{P_n^{i_{n-k+1}\ldots i_n i_1 \ldots i_{n-k}}}.
$$
In such a way, the derivative with respect to ${\bf v}$ becomes 
$$
\partial_{v_l} F_f=\sum_{n=0}^\infty n F_{P_n^{(i_1 \ldots i_n)}} ~ v_{i_1}\cdots v_{i_{n-1}} \delta_{i_n}^l .
$$
In these variables, the Hamiltonian becomes 
\begin{equation}
H(\{P_n\},{\mathbf{E}},{\mathbf{B}}) =
		\int d^3x~ \frac{P_2^{ii}}{2}
		+ \int d^3x~ \frac{{\mathbf{E}}^2+{\mathbf{B}}^2}{2},
\label{HamiMome}
\end{equation}
and the expression of the bracket becomes
\begin{eqnarray}
\{F,G\}_V &=&
		 \sum_{m,n} \int d^3x~ m P_{n+m-1}^{\alpha \beta} \partial_k F_{P_n^{\alpha}} G_{P_m^{(\beta k)}} 
		+ \sum_{m,n} \int d^3x~nm P_{n+m-2}^{\alpha \beta} \frac{eB_{ij}}{2} F_{P_n^{(\alpha i)}} G_{P_m^{(\beta j)}}
		\nonumber \\
		&& + \sum_{n} \int d^3x~  e n P_{n-1}^{\alpha} F_{P_n^{(\alpha i)}} G_{E_i} + \int d^3x~   F_{\mathbf{E}} \cdot {\mathbf{\nabla}} \times G_{\mathbf{B}} -(F \leftrightarrow G), 
\label{BracMome}
\end{eqnarray}
where $B_{ij}= \varepsilon_{ijk}B^k$ with $ \varepsilon_{ijk}$ the Levi-Civita tensor, and $(F\leftrightarrow G)$ indicates that the terms obtained by inverting $F$ and $G$ in the summation have to be subtracted (in order to fulfill the antisymmetry property of the Poisson bracket). Here and in what follows, $\partial_k$ designates the partial derivative with respect to $x_k$; it acts on the next term, e.g. $\partial_k fg =(\partial_k f)g$. 
The greek indices $\alpha$ and $\beta$ in Eq.~(\ref{BracMome}) denote a set of indices so as to complete the summation. For instance, in $P_n^{(\alpha i)}$, the indices $\alpha$ is a set of $n-1$ indices $\alpha= (i_1,\ldots,i_{n-1})$ so that $P_n^{(\alpha i)}=P_n^{(i_1 \ldots i_{n-1} i)}$.

The commonly used fluid model corresponds to a truncation of the moments at order one, keeping as field variables only the fluid density $\rho=P_0$ and the momentum density ${\mathbf{M}}= P_1$  (or equivalently the fluid velocity defined as ${\mathbf{M}} / \rho$). 
The bracket~(\ref{BracMome}) has the particular property that the set of all functionals of the reduced field variables $(\rho,{\mathbf{M}},{\mathbf{E}},{\mathbf{B}})$ is invariant, in the sense that the Poisson bracket~(\ref{BracMome}) of two functionals $F(\rho,{\mathbf{M}},{\mathbf{E}},{\mathbf{B}})$ and $G(\rho,{\mathbf{M}},{\mathbf{E}},{\mathbf{B}})$ is again a functional of $(\rho,{\mathbf{M}},{\mathbf{E}},{\mathbf{B}})$. For this subset of functionals, the bracket reduces to
\begin{eqnarray}
	\{F,G\}_1 &=&
		\int d^3x~ \left( (P_{0}\partial_k F_{P_0}+P_1^j \partial_k F_{P_1^j})  G_{P_1^{k}} + e P_{0} \left( B_{ij} F_{P_1^{i}} G_{P_1^{j}}/2 +  F_{P_1^{i}} G_{E_i}\right)
		+ F_{\mathbf{E}} \cdot {\mathbf{\nabla}} \times G_{\mathbf{B}}\right) \nonumber \\
		&&  - (F\leftrightarrow G) ,
\label{BracFluid0}
\end{eqnarray}
which is indeed again a functional of $(\rho,{\mathbf{M}},{\mathbf{E}},{\mathbf{B}})$.
At this stage, the reduction of the dynamics has not yet been performed because the Hamiltonian $H$ given by Eq.~(\ref{HamiMome}) depends on $P_2$ so it does not belong to the subalgebra of functionals of $(\rho,{\mathbf{M}},{\mathbf{E}},{\mathbf{B}})$. In order to perform the reduction, one needs to make an assumption on the Hamiltonian. For example, in the so-called cold plasma model~\cite{Bitt04}, the reduced Hamiltonian is $H^* = \int d^3x~ ({\mathbf{M}}^2/(2\rho)+({\mathbf{E}}^2+{\mathbf{B}}^2)/2)$, i.e., the original kinetic part of the Hamiltonian $\int d^3x~  P_2^{ii}/2$ has been replaced by $\int d^3x~ {\mathbf{M}}^2/(2\rho)$. Practically, the choice is inspired by the physics of the system under consideration. In the presence of scalar pressure terms and in the absence of heat flux (adiabatic closure), the entropy $s$ has to be included following Ref.~\cite{Bitt04}.
To account for this, $s$ has to be considered as an independent scalar field advected with the fluid. This is achieved by including the advection term given by
$$
		 \int d^3 x ~ G_{s} F_{{\mathbf{M}}} \cdot {\mathbf{\nabla}} s - (F\leftrightarrow G),
$$
into the bracket. The addition of these terms preserves the Jacobi identity, since it only corresponds to a passive scalar advected by the Hamiltonian fluid. The last step is to include the pressure term in the Hamiltonian which may be any function $\rho U(\rho,s)$. Accordingly, $U(\rho,s)$ stands for the thermodynamical internal energy per unit mass, and the pressure is $\rho^2 \partial_\rho U$. The model we get is the most common Hamiltonian fluid model (see Ref.~\cite{Morr81}) with the Hamiltonian 
$$
H[\rho,{\mathbf{M}},{\mathbf{E}},{\mathbf{B}}] =\int d^3 x~\left(\frac{{\mathbf{M}}^2}{2\rho}+\rho U(\rho,s)+ \frac{{\mathbf{E}}^2+{\mathbf{B}}^2}{2}\right),
$$
and the bracket
\begin{eqnarray*}
\{F,G\} &=&
		\int d^3x~ \Big[ \rho ~ G_{{\mathbf{M}}} \cdot {\mathbf{\nabla}} F_{\rho}
		+ {\mathbf{M}} \cdot ~ (G_{{\mathbf{M}}} \cdot {\mathbf{\nabla}}) F_{{\mathbf{M}}} + G_{s} ~ F_{{\mathbf{M}}} \cdot {\mathbf{\nabla}} ~ s
		+ e\rho {\mathbf{B}} \cdot F_{{\mathbf{M}}} \times G_{{\mathbf{M}}}/2 \\
		&&\quad \qquad + e\rho F_{{\mathbf{M}}} \cdot G_{{\mathbf{E}}}
		+ F_{\mathbf{E}} \cdot {\mathbf{\nabla}} \times G_{\mathbf{B}}\Big]  - (F\leftrightarrow G) .
\end{eqnarray*}
The introduction of $U$ in the Hamiltonian is a way of keeping some information that is enclosed in the pressure tensor. As a result, this model is almost a reduction at order $1$ in the moment series, but it is related to a closure at order $2$ as well.

\section{Fluid reduction at orders higher than 1}
\label{sec:3}

As it was shown above, the Hamiltonian fluid reduction at order $1$ is given by a Poisson subalgebra of the parent Hamiltonian structure. This results in a reduced bracket that is just a truncation of the initial bracket~(\ref{BracMome}). However, in many cases such a fluid model with only the first two moments is not rich enough to account for all the physics of interest, and one has to include kinetic effects and retain higher order moments in the fluid model. In this section we consider the Hamiltonian derivation of such higher order models. It should be pointed out that including the second order moments would be particularly interesting since the Hamiltonian belongs to the reduced algebra, and the closure does not affect the Hamiltonian.

The goal is to consider generalizing the results of the previous section to get fluid models of order higher than $1$ by truncating the initial bracket. This method seems natural in the case of the fluid reduction, which relies on the idea that all the physical information is contained in the set $\{P_n\}_{n\leqslant N}$ of the $N$ first moments of $f$. The remaining moments will have to be expressed as functionals of them, $P_{i>N}=\Phi_i(\{P_n\}_{n\leqslant N})$. Here we consider the simplest closure assumption $P_n =0$ for $n >N$. 

It should be noticed that it would actually be best suited to consider moments defined by quantities like $\Pi_n = \int d^3~ f ({\mathbf{v}}- {\bf M} / {\rho})^{\otimes n}/ \rho$. However it makes the discussion slightly more complicated, so for the sake of clarity, we have kept the moments as defined by Eq.~(\ref{DefMom}) in this section. We consider reduced brackets defined from the moments $\Pi_n$ in Sec.~\ref{sec:4}. 

We look for the reduced bracket by removing the undesired moments from the initial Poisson bracket~(\ref{BracMome}), i.e., we consider the sub-algebra of functionals $F(f,{\bf E},{\bf B})=\bar{F}(\{P_n\}_{n\leq N},{\bf E},{\bf B})$ and we truncate the Poisson bracket~(\ref{BracMome}) acting on two functionals of this sub-algebra by removing all terms proportional to $P_n$ for $n> N$. This bracket does not involve functional derivatives $F_{P_n}$ for $n> N$ since these terms vanish when acting of an element of the subalgebra. We denote this truncated bracket $\{F,G\}_N$. Since $\{F,G\}_N=\{F,G\}_V$ for all functionals $F$ and $G$ of this sub-algebra by enforcing $P_n=0$ for $n> N$, the reduced bracket is automatically algebraically closed and it retains from the initial bracket $\{\cdot,\cdot\}_V$ the bilinearity, the antisymmetry and the Leibniz rule. The only property one has to check to have a Hamiltonian structure is the Jacobi identity, since this property does not automatically transfer to truncated brackets. 

In order to be more specific, we inspect more closely the case $N=2$. The reduced bracket is given by
$$
\{F,G\}_2=\{F,G\}_1+\{F,G\}'_2,
$$
where the bracket $\{F,G\}_1$ is given by Eq.~(\ref{BracFluid0}) and $\{F,G\}'_2$ is a bracket which only involves $F_{P_2}$ or $G_{P_2}$. This bracket is given by
\begin{eqnarray*}
\{F,G\}'_2&=& \int d^3x~ \left( 2(P_1^j \partial_k F_{P_0} +P_2^{ij}\partial_k F_{P_1^i})G_{P_2^{(jk)}} + P_2^{ij}\partial_k F_{P_2^{ij}} G_{P_1^k}+2eP_1^i F_{P_2^{(ij)}} G_{E_j}\right. \\
&& \quad \qquad \left. +2eB_{ij}(P_1^k F_{P_2^{(ik)}} G_{P_1^j}+P_2^{kl} F_{P_2^{(ik)}} G_{P_2^{(jl)}})\right) - (F\leftrightarrow G).  
\end{eqnarray*}
The bracket given by Eq.~(\ref{BracMome}) between functionals of $(\rho,{\mathbf{M}}, P_2, {\mathbf{E}}, {\mathbf{B}})$ involves only one term that is not a functional of $(\rho,{\mathbf{M}}, P_2, {\mathbf{E}}, {\mathbf{B}})$, and which is proportional to $P_3$, and it is given by
$$
\{F,G\}_2''=2 \int d^3x~ P_3^{ijk} \partial_l F_{P_2^{ij}} G_{P_2^{(kl)}} - (F\leftrightarrow G).
$$  
The reduction involves $P_3=0$, it is why this term has been dropped from the reduced bracket $\{\cdot,\cdot\}_2$. However we can not conclude that the truncated bracket $\{\cdot ,\cdot \}_2$ is Hamiltonian since the Jacobi identity has to be checked a posteriori.
We show that in fact some terms not proportional to $P_3$ are generated in the Jacobi identity by the contribution $\{\cdot,\cdot \}_2''$. These contributions originate from the terms in the bracket~(\ref{BracMome}) that are proportional to $P_n$ for $n\leq 2$ and which involve $F_{P_3}$ or $G_{P_3}$. These terms are 
$$
\{F,G\}_3^c= 3\int d^3x~\left( P_2^{ij} \partial_k F_{P_0} G_{P_3^{(ijk)}}+eB_{ij} P_2^{kl} F_{P_1^i} G_{P_3^{(jkl)}}+ eP_2^{ij} F_{P_3^{(ijk)}} G_{E_k}\right)- (F\leftrightarrow G)
$$
If we restrict this additional contributions to the first term, i.e., in absence of magnetic and electric field, the non-trivial contribution in the Jacobi identity comes from the bracket $\{\{F,G\}_2'',H\}_3^c$ which includes terms of the form 
$$
-2\int d^3x~ P_2^{ij}\partial_l F_{P_2^{ij}} G_{P_2^{(kl)}} \partial_k H_{P_0},
$$
and as well two other terms, that have the same expression, but with circular permutation of $(ijk)$. 
This contribution to the Jacobi identity of the bracket~(\ref{BracMome}) suggests a counter-example for the failure of the Jacobi identity for the truncated bracket $\{\cdot,\cdot \}_2$. We select $F$ as a functional of $P_2$, $G$ as a functional of $P_2$ and $H$ as a functional of $P_0$. For instance, we choose $F=\int d^3x~(P_2^{11})^2/2$, $G=\int d^3x~P_2^{22}$ and $H=P_0$ as a counterexample of the Jacobi identity. It leads to 
$$
 \{\{F,G\}_2,H\}_2+\circlearrowleft = -\partial_2^2\left( (P_2^{11})^2\right)-4\partial_1\left(P_2^{12}\partial_2 P_2^{11}\right). 
$$
The same reasoning can be performed at order $N$. The number of families of possible counter-examples to the Jacobi identity increases as $N$ increases. However a convenient one is inspired from the case $N=2$ and involves a functional of $P_N$, a functional of $P_2$ and a functional of $P_0$. It follows that the truncated bracket $\{\cdot,\cdot\}_N$ neglects terms proportional to $P_{N+1}$ which provides a non-vanishing contribution to the Jacobi identity for $\{\cdot ,\cdot\}_N$ even in the subalgebra of functionals of $\{P_n\}_{n\leq N}$, ${\bf E}$ and ${\bf B}$. 

Therefore, starting from $N=2$, the truncated brackets $\{\cdot,\cdot\}_N$ do not satisfy the Jacobi identity. 

We notice that the reduced bracket $\{\cdot,\cdot\}_N$ contains most of the terms of the initial bracket $\{\cdot,\cdot\}_V$ when acting on functionals of the reduced variables $\{P_n\}_{n\leq N}$, ${\bf E}$ and ${\bf B}$. It is then more efficient to study the few removed terms than the many kept ones. This suggests a simpler way of verifying the Jacobi identity by only considering the truncated terms. It is explained in the Appendix.

\section{Brackets expressed using pressure-like moments}
\label{sec:4}

In Refs.~\cite{Shad04,Shad05,Shad11}, a warm fluid model for a collisionless low temperature relativistic plasma has been introduced. It gives dynamical equations for moments of the kinetic (Vlasov) distribution up to order two, i.e., it involves $P_0$ (fluid density), $\Pi_1=P_1/P_0$ (fluid velocity) and 
$$ 
\Pi_2^{ij}=\frac{1}{P_0}\int d^3v~(v_i-\Pi_1^i)(v_j-\Pi_1^j)  f.
$$ 
It is argued in Refs.~\cite{Shad04,Shad11} that this model is Hamiltonian and the bracket is given by
\begin{equation}
\label{eq:Shad}
\{F,G\}_2=\{F,G\}_1+\{F,G\}_2',
\end{equation}
where 
$$
\{F,G\}_1=\int d^3x~\left( \partial_k F_{P_0} G_{\Pi_1^k}+\frac{1}{2P_0}(\partial_k \Pi_1^l-\partial_l \Pi_1^k +e B_{kl}) F_{\Pi_1^{k}} G_{\Pi_1^{l}} + e F_{\Pi_1^k} G_{E_k}+F_{\bf E}\cdot \nabla \times G_{\bf B}\right)-(F \leftrightarrow G).
$$
and
\begin{eqnarray*}
\{F,G\}'_2&=&
\int d^3x~ \left(\frac{\partial_k \Pi_2^{rs}}{P_0} F_{\Pi_1^k}G_{\Pi_2^{(rs)}}+ 2\Pi_2^{rs} \partial_k \left(\frac{F_{\Pi_1^s}}{P_0}\right) G_{\Pi_2^{(rk)}}\right.\\
&&\qquad \left. + \frac{2\Pi_2^{rs}}{P_0}(\partial_k\Pi_1^l-\partial_l \Pi_1^k + eB_{kl})F_{\Pi_2^{(kr)}}G_{\Pi_2^{(ls)}}\right) -(F \leftrightarrow G) .
\end{eqnarray*}
This model is derived in a very similar way as in the previous section, except that, instead of going from $f$ to $P_n$, the change of variables is from $f$ to $\Pi_n$ where $\Pi_0=P_0$, $\Pi_1=P_1/P_0$, and for $n\geqslant 2$, $\Pi_n=\int d^3v~ ({\bf v}-\Pi_1)^{\otimes n}/P_0$. The reduction at order $N$ corresponds to setting $\Pi_n =0$ for $n >N$. In particular, in the derivation of the model given in Ref.~\cite{Shad04}, the contributions in the bracket which are proportional to $\Pi_3$ have been neglected given a specific assumption on the Vlasov distribution. However, as in the previous section, these neglected terms contribute to satisfying the Jacobi identity. Given the counter-example 
\begin{eqnarray*}
&& F= P_0 \Pi_2^{11},\\
&& G=\int d^3x~P_0 \Pi_2^{22},\\
&& H=\int d^3x~P_0 \Pi_2^{33}, 
\end{eqnarray*}
the bracket(\ref{eq:Shad}) fails to satisfy the Jacobi identity since
$$
\{\{F,G\},H\}+\circlearrowleft =  8P_0\Pi_2^{12}\partial_2\left(\frac{1}{P_0}\partial_3(P_0 \Pi_2^{13})\right)+4\partial_2 \Pi_2^{11} \partial_3 (P_0 \Pi_2^{23})-(2\leftrightarrow 3),
$$
where $(2\leftrightarrow 3)$ indicates the same two terms where $\Pi_2^{12}$ has been exchanged with $\Pi_2^{13}$, and $\partial_2$ with $\partial_3$.

\section{Conclusion}

In Ref.~\cite{Shad04}, it is mentioned that the reduction ``is exact: for any functionals $F[n,{\bf P},{\bm \Pi}]$ and $G[n,{\bf P},{\bm \Pi}]$ we have $\{F,G\}_M=\{F,G\}_V$. As a consequence we see that the moment bracket inherits the Jacobi identity (as well as all other properties) from the full bracket.'' Here the bracket $\{\cdot,\cdot\}_M$ refers to the reduced bracket $\{\cdot,\cdot\}_2$ given by Eq.~(\ref{eq:Shad}) and $\{\cdot,\cdot\}_V$ refers to the Vlasov-Maxwell bracket~(\ref{VlasMaxw}). The first part of the sentence is correct. This is why the reduced bracket inherits all the properties of the Vlasov-Maxwell bracket that are linked to the values of the bracket: bilinearity, antisymmetry and Leibnitz rule. However the Jacobi identity involves not only the values of the bracket but also their gradients, and in general, 
$$
\{\{F,G\}_M,H\}_M\not= \{\{F,G\}_V,H\}_V,
$$ 
even if $\{F,G\}_M=\{F,G\}_V$ (after some assumption on the Vlasov distribution which is not preserved by the flow, or after neglecting higher order terms), so the reduced bracket can fail the Jacobi identity. 

Here we have shown that deriving reduced fluid models from kinetic models like Vlasov-Maxwell equations has to be handled with care as soon as moments of order two or higher are considered. As an illustration we have shown that the model introduced in Ref.~\cite{Shad04} involving moments up to order two does not satisfy the Jacobi identity, by exhibiting a counter-example.

\section*{Acknowledgments}

 We acknowledge financial support from the Agence Nationale de la Recherche (ANR GYPSI). This work was also supported by the European Community under the contract of Association between EURATOM, CEA, and the French Research Federation for fusion study. The views and opinions expressed herein do not necessarily reflect those of the European Commission.  We would like to acknowledge several helpful conversations with the \'Equipe de Dynamique Nonlin\'eaire of the Centre de Physique Th\'eorique of Marseille.

\section*{Appendix: Verifying the Jacobi identity for truncated brackets}

First we consider a Poisson bracket written as
$$
\{F,G\}=F_{z_i} {\mathbb J}^{z_i z_j} G_{z_j}.
$$
The Poisson matrix ${\mathbb J}$ satisfies the Jacobi identity rewritten as
$\sum_{(ijk)} {\mathbb J}^{z_i z_l}\partial_{z_l}{\mathbb J}^{z_j z_k} =0$,
where $\sum_{(ijk)}$ means circular permutation of the indices $i, j, k$. For the reduction we split the set of variables in two subsets ${\bf z}=({\bf Z}, {\bm \zeta})$. The reduction is chosen as acting on the subalgebra of functions of ${\bf Z}$. In order to do that, we decompose the Poisson matrix into
$$
	{\mathbb J} = {\mathbb J}_0^{{\bf Z}{\bf Z}}+{\mathbb J}_1^{{\bf Z}{\bf Z}}+{\mathbb J}_R^{{\bf Z}{\bf Z}}
	+ {\mathbb J}_0^{[{\bm\zeta}{\bf Z}]}+{\mathbb J}_R^{[{\bm\zeta}{\bf Z}]}
	+ {\mathbb J}_0^{{\bf Z}{\bf Z}}+{\mathbb J}_R^{{\bf Z}{\bf Z}}  \,,
$$
where $[{\bm\zeta}{\bf z}]$ means antisymmetrization, i.e., ${\mathbb J}^{[{\bm\zeta}{\bf z}]}={\mathbb J}^{{\bm\zeta}{\bf z}}-{\mathbb J}^{{\bf z}{\bm\zeta}}$, the index in ${\mathbb J}_i$ indicates the $i$-th order in ${\bm\zeta}$, and ${\mathbb J}_R$ includes the remaining (higher order) terms. The reduced bracket is simply the truncation of ${\mathbb J}$ obtained by removing terms involving $\partial_{\bm\zeta}$ and taking the limit ${\bm \zeta}\rightarrow 0$. As a result,
$$
	{\mathbb J}_* = {\mathbb J}_0^{{\bf Z}{\bf Z}}.
$$
This matrix ${\mathbb J}_*$ is associated with a Hamiltonian system if and only if
$$
	\sum_{(ijk)} {\mathbb J}_0^{Z_i\zeta_l}\partial_{\zeta_l}{\mathbb J}_1^{Z_j Z_k} =0 \,.
$$
In the example of Sec.~\ref{sec:3}, ${\mathbb J}_0^{{\bf Z}{\bm\zeta}}$ corresponds to $\{\cdot ,\cdot\}_3^c$, and ${\mathbb J}_1^{{\bf Z}{\bf Z}}$ to $\{\cdot,\cdot\}_2''$.  
We notice that a Lie subalgebra corresponds to the case ${\mathbb J}_1^{{\bf Z}{\bf Z}}=0$ and ${\mathbb J}_R^{{\bf Z}{\bf Z}}=0$. The condition on the validity of the reduction can be generalized to the case ${\mathbb J}_1^{{\bf Z}{\bf Z}}=0$ which is sufficient to verify the Jacobi identity. However this assumption is not often verified since brackets generally contain linear terms in the variables~\cite{Zakh97}.


\end{document}